\documentclass[12pt]{iopart}
\usepackage{epsfig} 
\begin{document}

\title[Strange particle production in 158 and 40 $A$
GeV/$c$ Pb-Pb and p-Be collisions]{Strange particle 
production in 158 and 40 $A$ GeV/$c$ Pb-Pb and p-Be collisions} 

\author{D~Elia for the NA57 Collaboration\footnote[1]{For
the full author list see Appendix ``Collaborations'' in this
volume.}
}

\address{INFN Sezione di Bari, Bari, Italy} 

%

\begin{abstract}
Results on strange particle production in Pb-Pb 
collisions at 158 and 40 $A$ GeV/$c$ beam momentum
from the NA57 experiment at CERN SPS are presented.
Particle yields and ratios are compared with those measured
at RHIC. Strangeness enhancements with
respect to p-Be reactions at the same beam momenta
have been also measured: results about their dependence on centrality
and collision energy are reported and discussed.
\end{abstract}





\section{Introduction}

An enhanced production of multi-strange baryons and 
antibaryons in ultrarelativistic nucleus-nucleus collisions
with respect to proton-induced reactions
has been predicted long time ago as a signal of quark-gluon 
plasma formation~\cite{RafMue82-86}. 
Enhancements increasing with the strangeness content
of the particle have been first observed by WA97
at 158 $A$ GeV/$c$ beam momentum~\cite{And99}.
With the NA57 experiment those measurements
have been extended
over a wider centrality range and
to lower beam momentum~\cite{NA57prop}.
\\
The NA57 apparatus
was designed to detect strange and multi-strange
hyperons by reconstructing their weak decays
into final states containing only charged particles
($\Lambda$ $\rightarrow$ $\pi^-{p}$, 
\hspace{1mm} $\Xi^-$ $\rightarrow$ $\Lambda\pi^-$ and
$\Omega^-$ $\rightarrow$ $\Lambda{K}^-$ with $\Lambda$ $\rightarrow$ $\pi^-{p}$
and the corresponding decays for antiparticles).
Tracks are measured in the silicon telescope, a 30 cm long
array of pixel detector planes
with 5 $\times$ 5 cm$^2$ cross section:
the acceptance coverage corresponds to about half a unit of
rapidity at central rapidity and medium transverse momentum.
Additional pixel planes and
double-sided silicon microstrip detectors, placed
behind the telescope, were used as a lever arm detector
to improve the momentum resolution for high momentum tracks.
The centrality trigger, based on a scintillator
petal system placed 10 cm downstream of the target,
selected the most central 60\% of the inelastic
cross section for Pb-Pb collisions.
The centrality of the collision is estimated
from the charged particle 
multiplicity sampled at central rapidity by two stations 
of silicon strip detectors.
The apparatus was placed inside the
1.4 Tesla magnetic field of the GOLIATH magnet.
Further details can 
be found in~\cite{Man99}. In the following section
we give some details of the analysis procedure,
then in section 3 we show the dependence of the K$^0_{\rm S}$, $\Lambda$,
$\Xi$ and $\Omega$ yields on centrality and energy
for the Pb-Pb collisions. The NA57 yields are also
compared with results from the higher energy data on
Au-Au collisions at RHIC. 
Finally, the centrality dependence of
hyperon enhancements at both
energies are presented and discussed in section 4.

\section{Data samples and analysis}

The results on Pb-Pb collisions presented in this
paper are based on the full statistics 
collected at 158 and
40 $A$ GeV/$c$ beam momentum. 
Samples of p-Be collisions at low energy 
have been also collected,
while for the reference data at 158 $A$ GeV/$c$ 
the p-Be data from WA97 have been used.
\\
Clean particle signals 
with negligible residual background
have been obtained with geometrical 
and kinematical cuts: for details see \cite{Man2002}.
\\
The Pb-Pb sample has been divided into
five centrality classes.
As an estimate of the number of participant nucleons
we use the number
of wounded nucleons ($N_{wound}$) computed from 
the event multiplicity and the measured
trigger cross section via the Glauber model: for details on
the multiplicity analysis see \cite{Ant2000}.
Going from the most peripheral (class 0) to the most central
(class 4) bin, the average $N_{wound}$ for the Pb-Pb collisions
at 158 $A$ GeV/$c$ are
58 $\pm$ 4,
117 $\pm$ 4,
204 $\pm$ 3,
287 $\pm$ 2 and \linebreak
349 $\pm$ 1 respectively.
The most central class and the full 
range covered by all the five classes correspond
to 5\% and 53\% most central collisions respectively.
For each selected particle a weight is calculated by means of
a Monte Carlo procedure based on GEANT~\cite{GEANTref} used 
to estimate
geometrical acceptance and reconstruction efficiency losses.
\\
The double-differential distributions in rapidity $y$ and
transverse mass $m_{\rm T}$=$\sqrt{m^2+p_{\rm T}^2}$ 
for each particle type have been fitted
according to the following expression:

\begin{equation}
\frac{d^2N}{dm_T dy}=f(y) \hspace{1mm} m_T \exp\left(-\frac{m_T}{T_{app}}\right)
\hspace{65mm}.     
\label{eqmtfit}
\end{equation}

The inverse slope
parameters ($T_{app}$) have been extracted
using the maximum likelihood
method, assuming a flat rapidity distribution
within our acceptance region.
Results on the inverse slope parameters both in Pb-Pb 
and in p-Be collisions, together with a
study of the transverse mass spectra for Pb-Pb in
the framework of a blast wave model, are 
discussed in \cite{BrunoHQ04andPaper}.
\\
Particle yields have been
calculated as the number of particles per event extrapolated
to a common phase space
window, covering full $m_{\rm T}$ and one unit of rapidity
around midrapidity:

\begin{equation}
Y = \int_{m}^{\infty} dm_{\rm T} \int_{y_{cm}-0.5}^{y_{cm}+0.5} dy \frac{d^2N}{dm_{\rm T} dy}
\hspace{66mm}.     
\label{eqyield}
\end{equation}

The hyperon yields measured in p-Be collisions allow to
determine the strangeness enhancement factors. These
have been defined as

\begin{equation}
E = \left(\frac{Y}{<N_{wound}>}\right)_{Pb-Pb} / \left(\frac{Y}{<N_{wound}>}\right)_{p-Be}
\hspace{39mm}.     
\label{eqenhanc}
\end{equation}

\section{Energy and centrality dependence of the yields 
in Pb-Pb collisions}

\begin{figure}[htb]
\hspace{15mm} 
\includegraphics[scale=0.52]{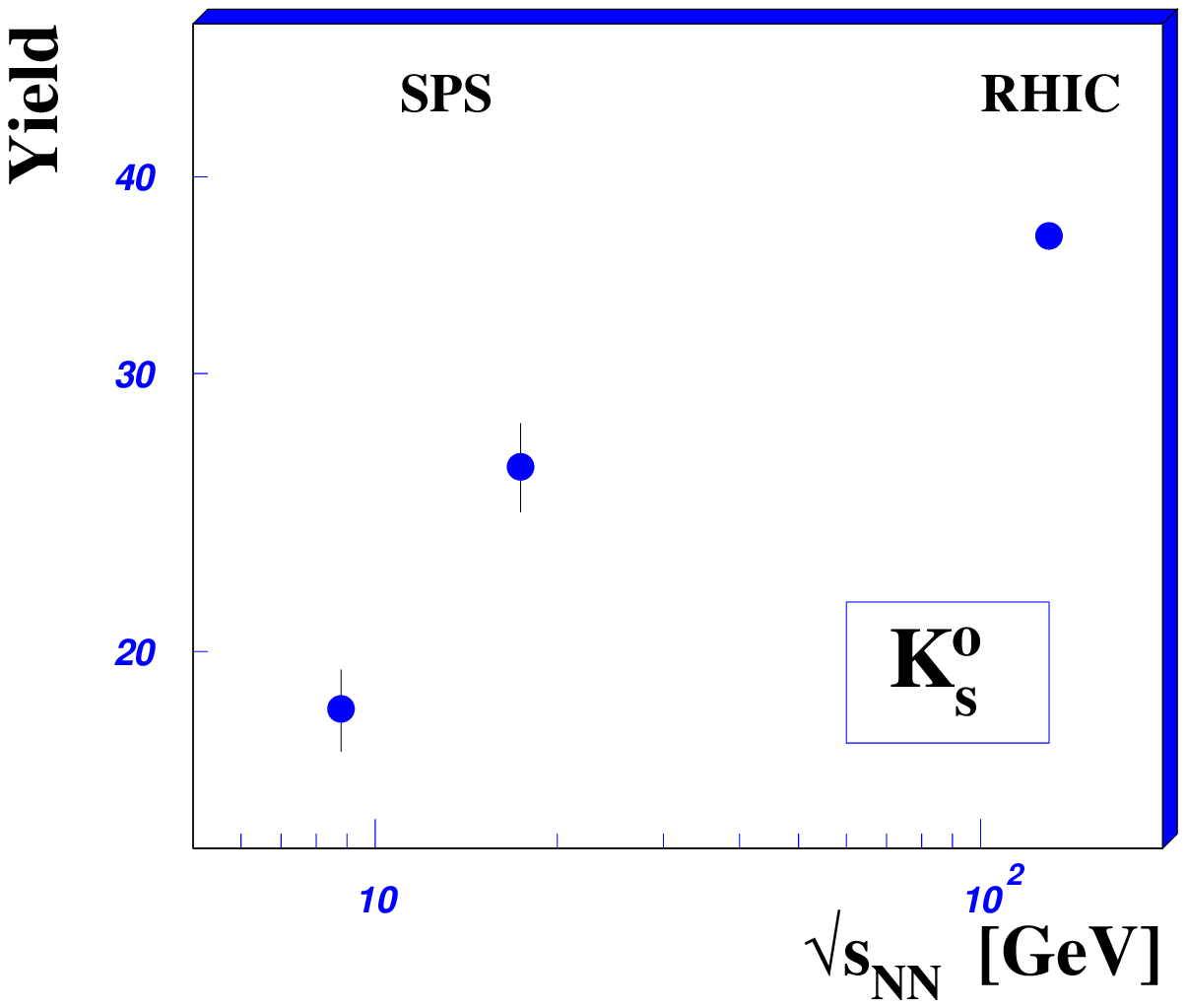}
\hspace{\fill}
\begin{minipage}[t]{110mm}
\includegraphics[scale=0.52]{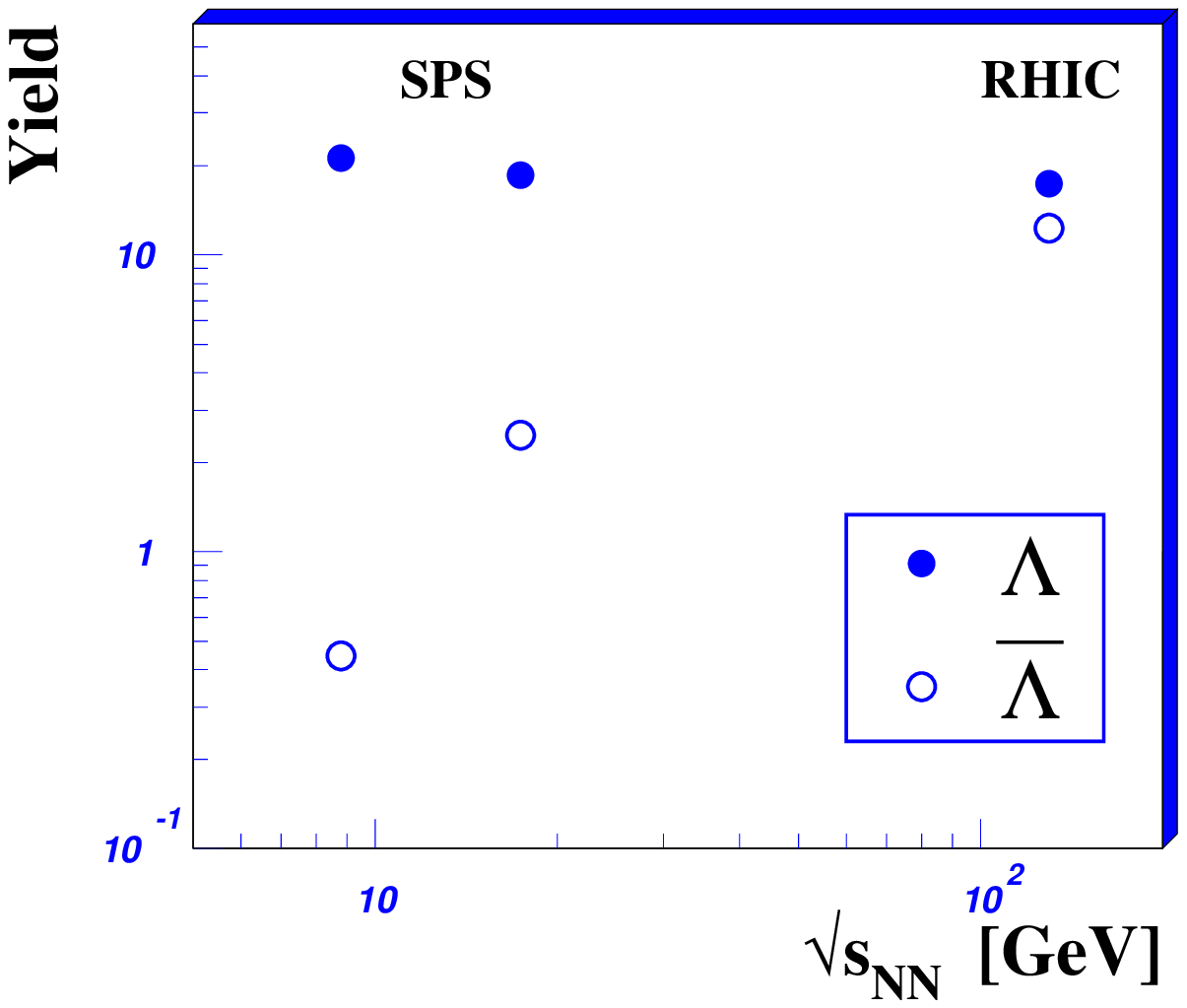}
\end{minipage}
\par
\hspace{15mm} 
\includegraphics[scale=0.52]{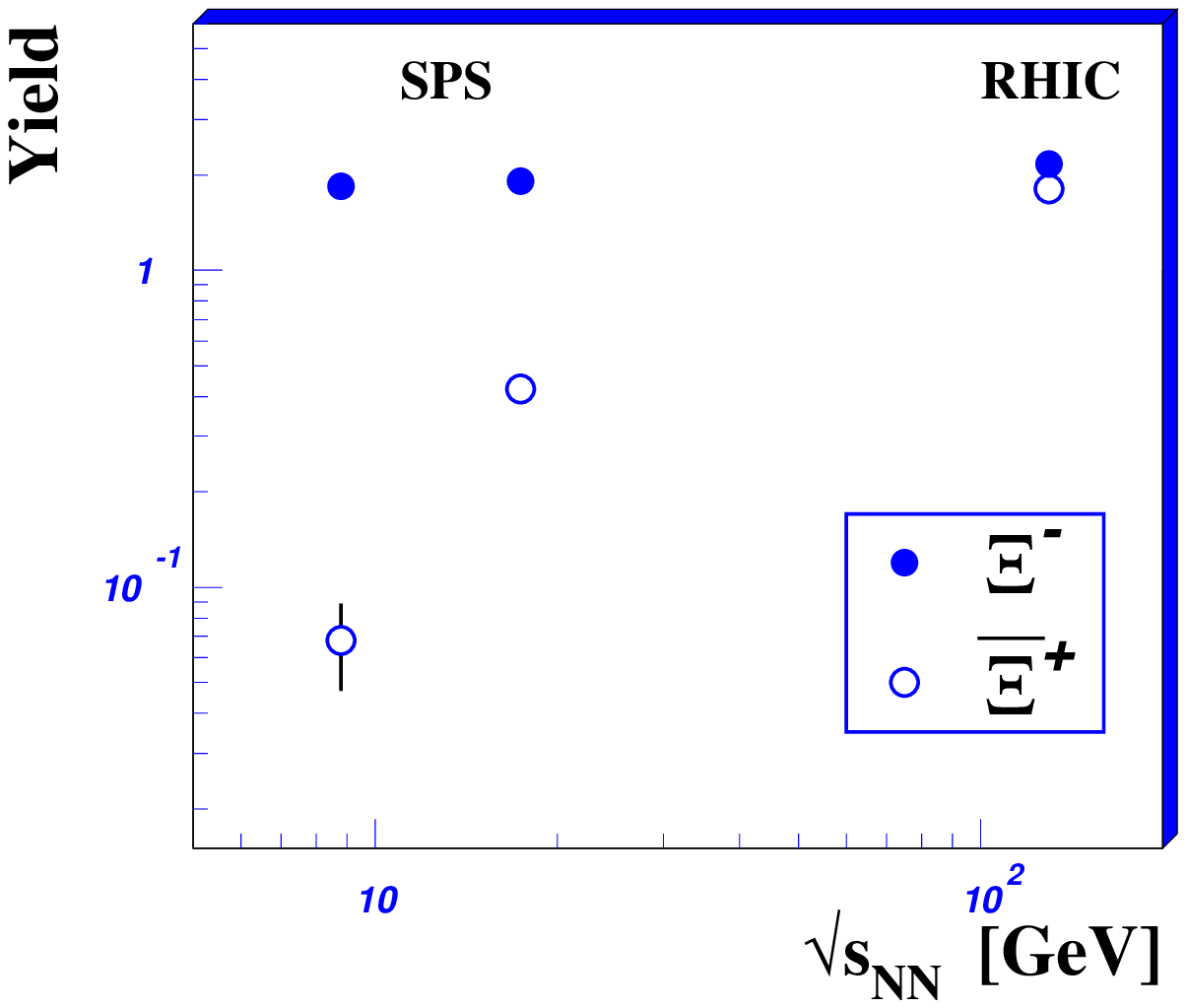}
\hspace{\fill}
\begin{minipage}[t]{110mm}
\includegraphics[scale=0.52]{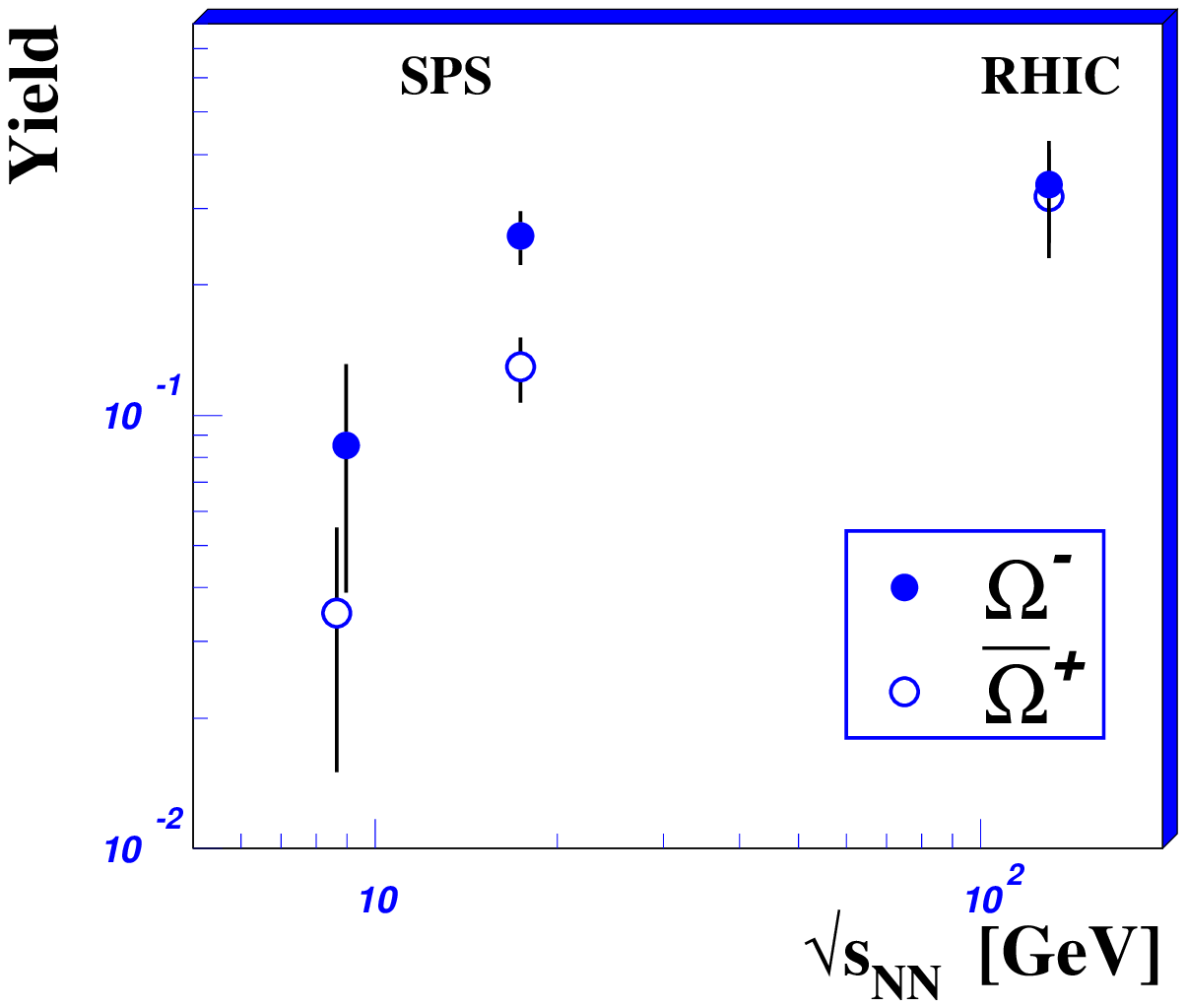}
\end{minipage}
\caption{\label{figyiecomp} K$^0_{\rm S}$ and hyperon yields
at central rapidity in nucleus-nucleus collisions at SPS and RHIC energies.
Large error bars on the $\Xi$ and $\Omega$ yields
are due to the restriction of the NA57 data sample
to the STAR centrality ranges.}
\end{figure}

K$^0_{\rm S}$ and hyperon yields in Pb-Pb collisions have been 
obtained as a function of the centrality
at both 158 and 40 $A$ GeV/$c$~\cite{EnerdepLett}.
The corresponding  
mid-rapidity yields have been measured at
higher energy by
the STAR Collaboration
for central Au-Au collisions at RHIC~\cite{STARyieldref}.
By restricting our data to the same
centrality ranges used in STAR
(i.e. most central 6\%, 5\%, 10\%, 11\%
collisions for K$^0_{\rm S}$, $\Lambda$, $\Xi$ and $\Omega$ respectively),
we can compare results over a wide energy range:
in Figure \ref{figyiecomp} our yields per
unit rapidity at 40 $A$ GeV/$c$
\linebreak ($\sqrt{s_{NN}}$ = 8.8 GeV) and 158 $A$ GeV/$c$
($\sqrt{s_{NN}}$ = 17.3 GeV) are shown togheter with those
from STAR at $\sqrt{s_{NN}}$ = 130 GeV.

\begin{figure}[h]
\centering
\includegraphics[scale=0.55]{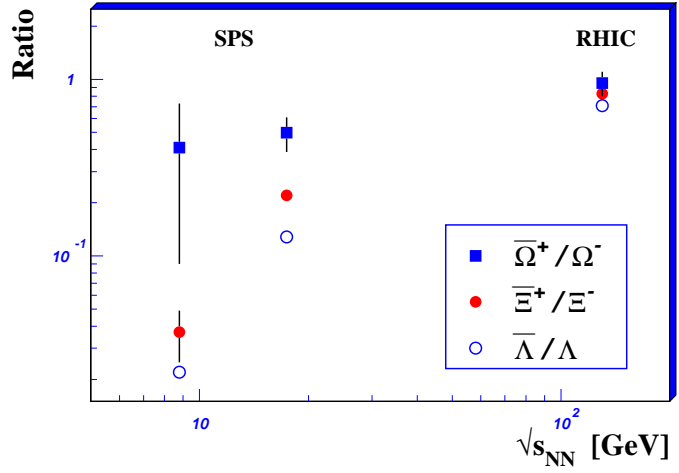}
\caption{\label{figratcomp} Antihyperon to hyperon
ratios in nucleus-nucleus collisions at SPS and RHIC.} 
\end{figure}

The $\Lambda$ and $\Xi^-$ yields
do not vary much from SPS to RHIC,
while a clear energy dependence
is observed for K$^0_{\rm S}$ and all three antihyperons. 
The antihyperon to hyperon ratios are plotted in
Figure \ref{figratcomp} as a function of
$\sqrt{s_{NN}}$ from SPS to RHIC~\cite{Adams2003PL}.
The ratios increase with increasing strangeness 
content of the hyperon, both at RHIC and SPS energies.
They also increase as a function
of the energy, the dependence being weaker for
particles with higher strangeness content. 
This can be understood as due to a
decrease of the baryon density at midrapidity with 
increasing collision energy.

\begin{figure}[htb]
\hspace{10mm}
\includegraphics[scale=0.56]{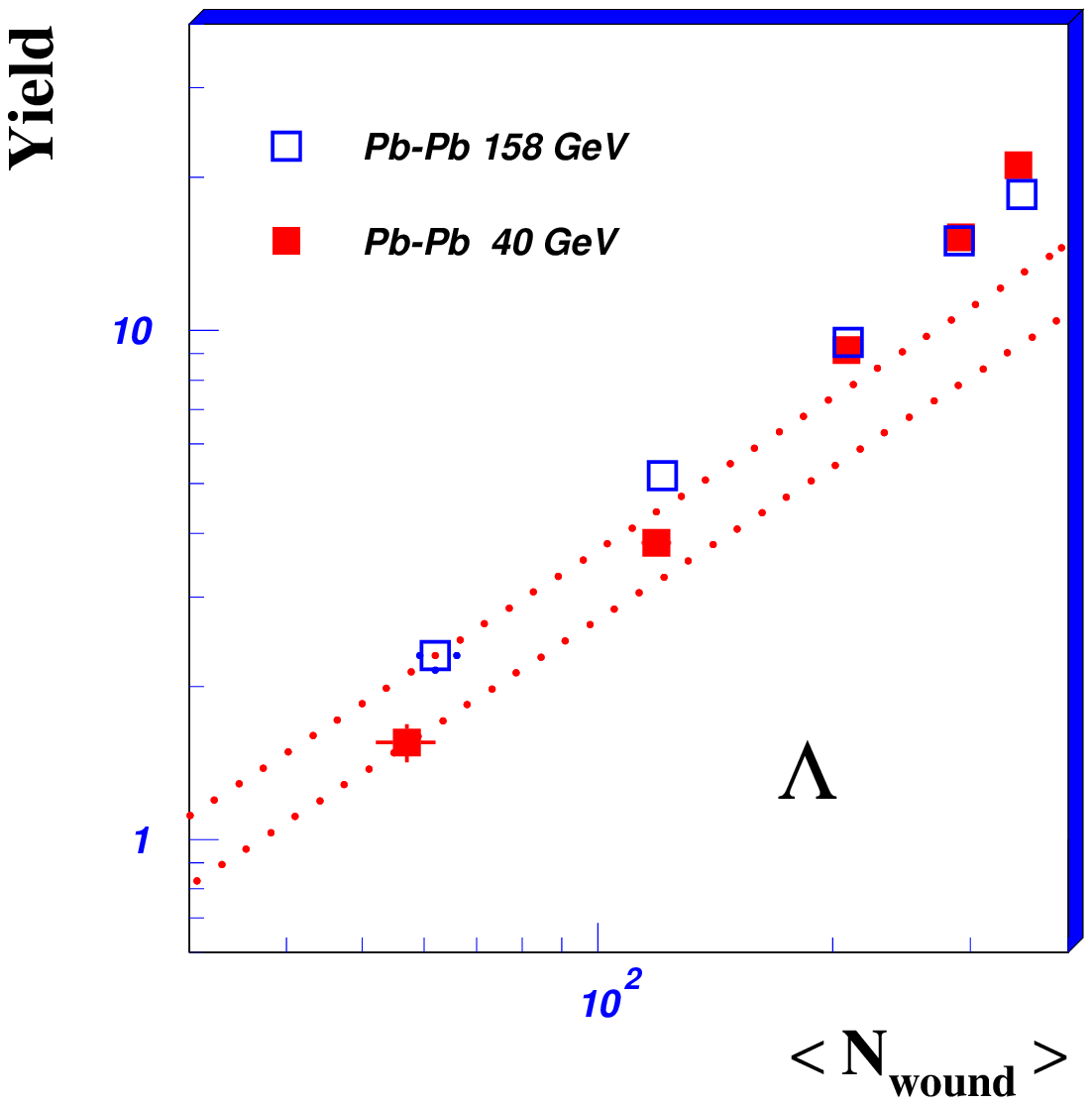}
\hspace{\fill}
\begin{minipage}[t]{110mm}
\includegraphics[scale=0.56]{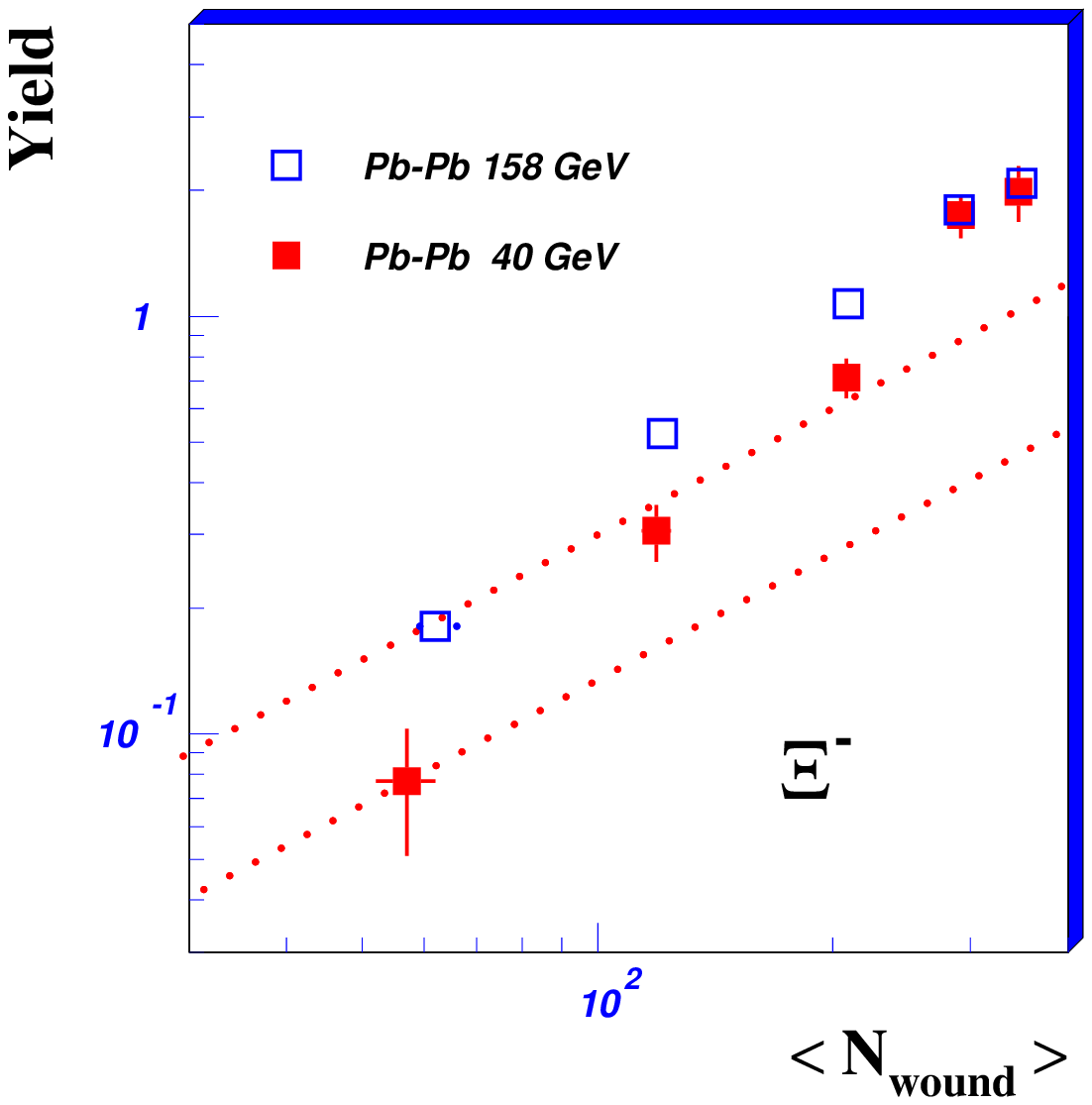}
\end{minipage}
\caption{\label{figyierap} Centrality dependence of $\Lambda$ and $\Xi^-$ 
yields in Pb-Pb collisions at SPS.} 
\end{figure}

The behaviour of the yields 
with the collision centrality 
has also been studied.
As an example, in Figure \ref{figyierap} 
the $\Lambda$ and $\Xi^-$ yields are shown for
each centrality class for both the 158 and 40 $A$ GeV/$c$
Pb-Pb collisions data samples. 

The dotted lines indicate the expected behaviour in case of
a linear increase of the yields with the number of
wounded nucleons. 
The $\Lambda$ and $\Xi^-$ yields
for central collisions at the two energies are very close.
The yields grow faster than linearly
with the number of participants, with a steeper centrality dependence 
for the lower energy data. 
Similar behaviour is
shown by K$^0_{\rm S}$ and $\overline\Lambda$, while 
the low statistics in the 40 $A$ GeV/$c$ sample does not allow
a firm conclusion for 
$\overline\Xi^+$ and $\Omega$.

\section{Strangeness enhancements with respect to p-Be collisions}

\begin{figure}[h]
\centering
\includegraphics[scale=0.49]{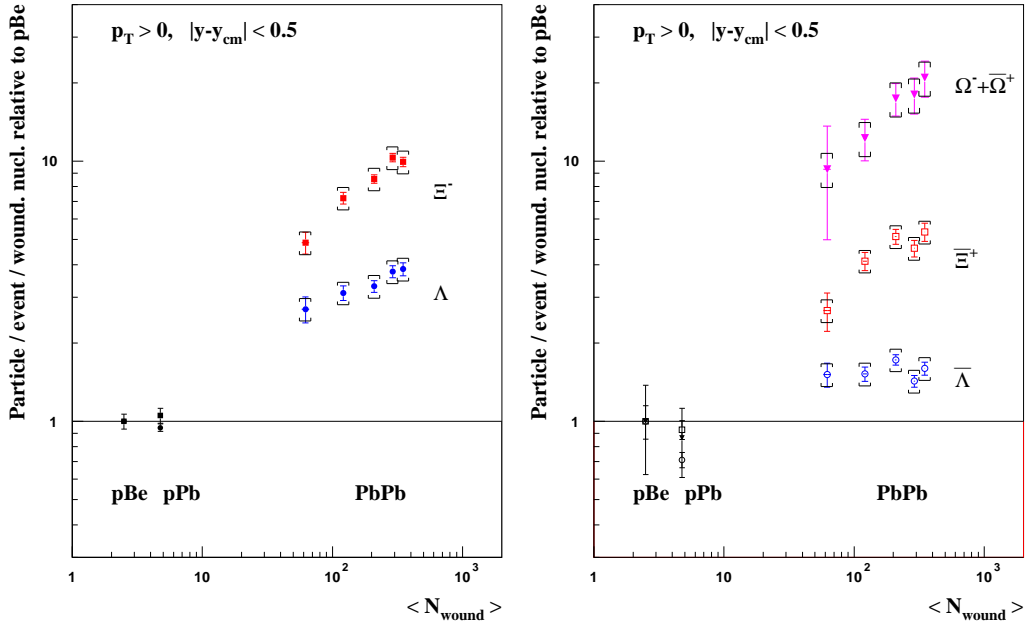}
\caption{\label{figyields_all} Centrality dependence of hyperon
enhancements at 158 $A$ GeV/$c$. 
The error bars indicate the statistical errors only,
while the bracket symbols represent the systematic errors.} 
\end{figure}

The results on strangeness enhancements at 158 $A$ GeV/$c$ are
shown in Figure \ref{figyields_all}.
Particles
containing at least one valence quark in common with the
nucleon (left) are kept separated from the others (right).
These results confirm the strangeness enhancement hierarchy
($E(\Lambda) < E(\Xi) < E(\Omega)$) already observed 
by WA97~\cite{And99}.

\begin{figure}[htb]
\centering
\includegraphics[scale=0.45]{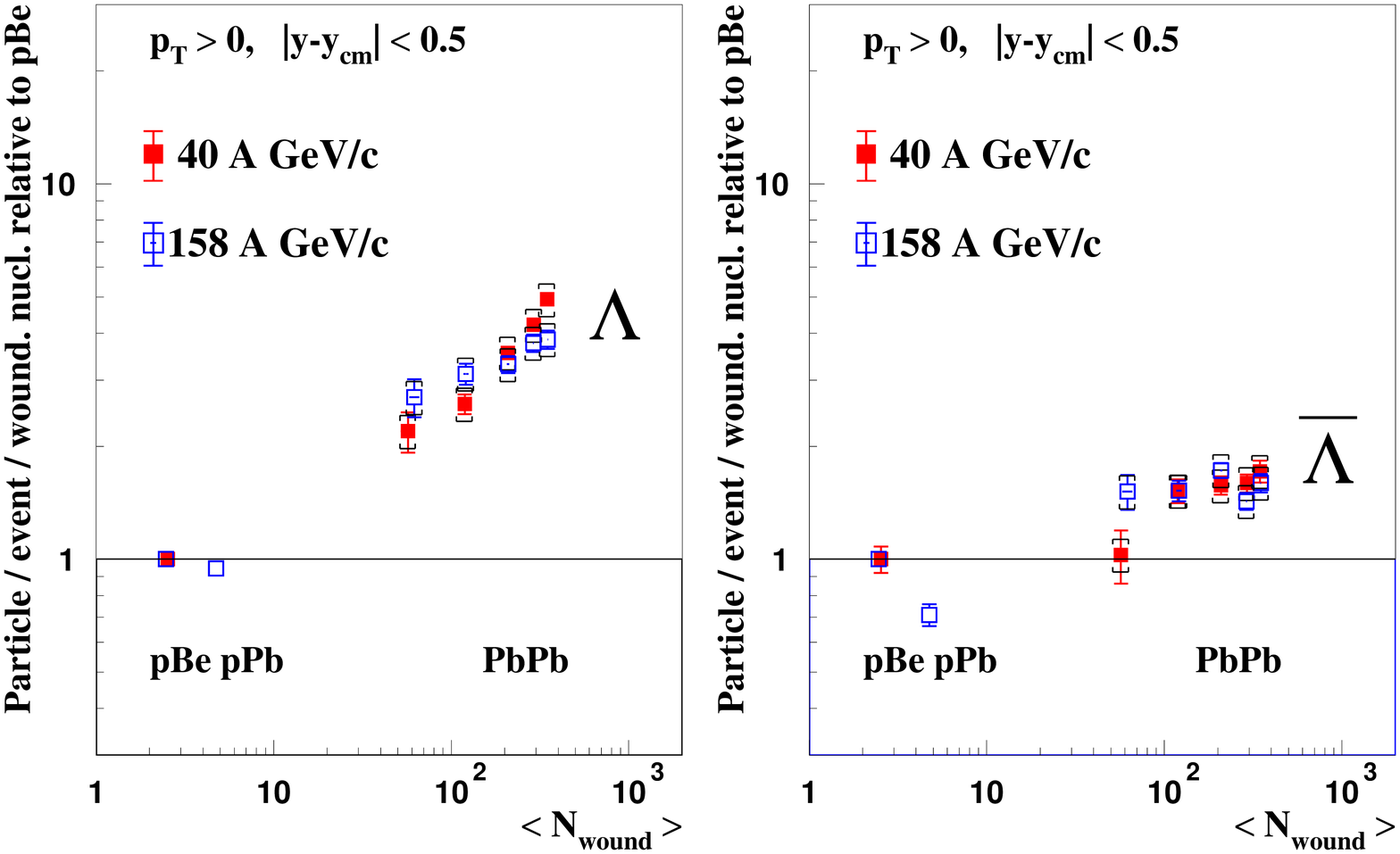}
\\
\begin{center}
\includegraphics[scale=0.45]{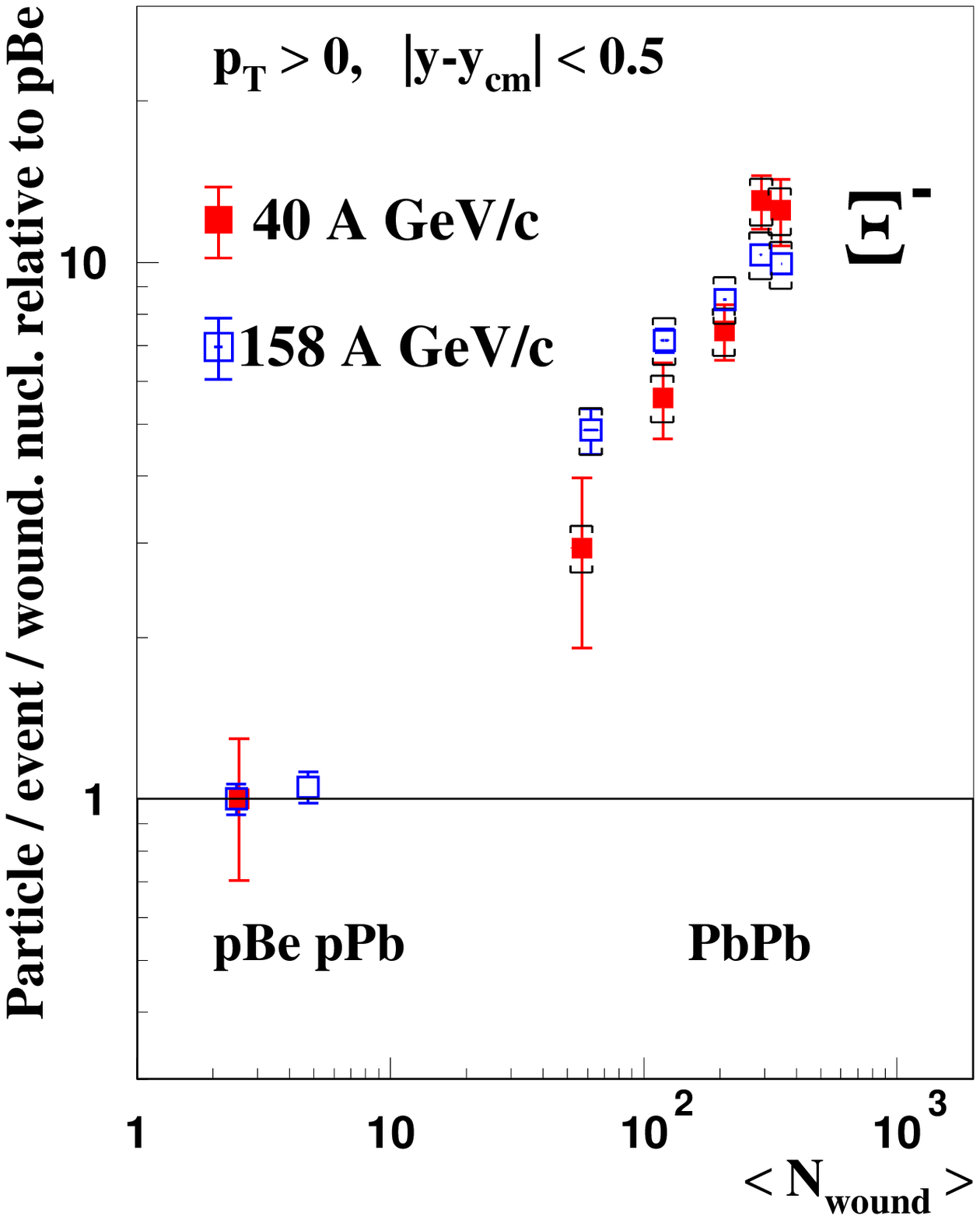}
\end{center}
\caption{\label{figyie_40} Centrality dependence of
$\Lambda$, $\overline\Lambda$ and $\Xi^-$
enhancements at both 158 and 40 $A$ GeV/$c$.
The error bars indicate the statistical errors only,
while the bracket symbols represent the systematic errors.} 
\end{figure}

We observe a significant centrality dependence
of the yields per wounded nucleon
for all hyperons except for $\overline\Lambda$.
\\
Similar enhancement factors
are observed also at lower energy: in Figure \ref{figyie_40}
the $\Lambda$, 
$\overline\Lambda$ and $\Xi^-$ enhancements measured
at both energies are shown
together
as a function of the collision centrality.
The low statistics in the p-Be reference sample allows only to estimate 
a lower limit for the $\overline\Xi^+$ enhancements at
40 $A$ GeV/$c$~\cite{GppeQM04}. The enhancement pattern
follows the same hierarchy observed at higher energy,
i.e. $E(\Lambda) < E(\Xi^-)$ and 
$E(\overline\Lambda) < E(\overline\Xi^+)$. 
As already seen for the yields, a steeper increase of the
enhancements with the number of participants
is observed for the lower energy sample. For the most central 
collisions (classes 3 and 4) the enhancements
are found to be larger at 40 than at 158 $A$ GeV/$c$ (by about 15-20\%).
\\
The strangeness enhancement, when described as a consequence of
the transition from the canonical to the asymptotic grand canonical 
limit, is indeed predicted to be a 
decreasing function of the collision energy~\cite{Tounsi}.
However, that model neither reproduces
the amount for central collisions nor the centrality
dependence of the measured enhancements.

\section{Conclusions} 

Results on strange particle production from
the NA57 experiment in 
158 and 40 $A$ GeV/$c$ Pb-Pb and p-Be collisions have been 
reported and discussed. 
\\
From $\sqrt{s_{NN}}$ = 8.8 GeV to
$\sqrt{s_{NN}}$ = 130 GeV 
the $\Lambda$ and $\Xi^-$ yields per unit
rapidity remain roughly constant. A clear 
increase of the yields with the collision energy is instead observed 
for all three
antihyperons. The antihyperon to hyperon ratios
increase with energy,
with a stronger dependence
for particles with lower strangeness content.
Such a pattern is indicative of
a decrease of the baryon density in the central region
as the energy is increased.
\\
The centrality dependence
of the yields and of the enhancements with respect to p-Be
is steeper at 40 than at 158 $A$ GeV/$c$. The enhancement 
pattern follows the same hierarchy with the strangeness
content as seen in the higher energy data,
with $E(\Lambda) < E(\Xi^-)$ and 
$E(\overline\Lambda) < E(\overline\Xi^+)$. 
For very central collisions (about 10\%) the 
enhancements are larger at 40 than at 158 $A$ GeV/$c$.

\end{document}